\documentclass[pdflatex,sn-mathphys-num]{sn-jnl}

\usepackage{graphicx}%
\usepackage{multirow}%
\usepackage{amsmath,amssymb,amsfonts}%
\usepackage{amsthm}%
\usepackage{mathrsfs}%
\usepackage[title]{appendix}%
\usepackage{xcolor}%
\usepackage{textcomp}%
\usepackage{manyfoot}%
\usepackage{booktabs}%
\usepackage{algorithm}%
\usepackage{algorithmicx}%
\usepackage{algpseudocode}%
\usepackage{listings}%
\usepackage{todonotes}
\usepackage{fullpage}
\usepackage{soul}
\DeclareUnicodeCharacter{0301}{\textvisiblespace}

\begin{document}
\title[Article Title]{Superconductivity in PrNiO$_2$ infinite-layer nickelates}

\author*[1]{\fnm{Hoshang} \sur{Sahib}}\email{hoshang.sahib@ipcms.unistra.fr}
\author[2]{\fnm{Francesco} \sur{Rosa}}
\author[3,4]{\fnm{Aravind} \sur{Raji}}
\author[2,5]{\fnm{Giacomo} \sur{Merzoni}}
\author[2]{\fnm{Giacomo} \sur{Ghiringhelli}}
\author[6]{\fnm{Marco} \sur{Salluzzo}}
\author[3]{\fnm{Alexandre} \sur{Gloter}}
\author[1]{\fnm{Nathalie} \sur{Viart}}
\author*[1]{\fnm{Daniele} \sur{Preziosi}}\email{daniele.preziosi@ipcms.unistra.fr}

\affil[1]{Universitè de Strasbourg, CNRS, IPCMS UMR 7504, F-67034 Strasbourg, France}
\affil[3]{Laboratoire de Physique des Solides, CNRS, Université Paris-Saclay, 91405 Orsay, France}
\affil[4]{Synchrotron SOLEIL, L’Orme des Merisiers, BP 48 St Aubin, 91192 Gif sur Yvette, France}
\affil[2]{Dipartimento di Fisica, Politecnico di Milano, Piazza Leonardo da Vinci 32, I-20133 Milano, Italy}
\affil[5]{European XFEL, Holzkoppel 4, Schenefeld, D-22869, Germany}
\affil[6]{CNR-SPIN Complesso di Monte S. Angelo, via Cinthia - I-80126 Napoli, Italy}

\abstract{
Several reports about infinite-layer nickelate thin films suggest that the superconducting critical temperature versus chemical doping phase diagram has a dome-like shape, similar to cuprates.
Here, we demonstrate a highly reproducible superconducting state in undoped PrNiO$_2$ thin films grown onto SrTiO$_3$.
Scanning transmission electron microscopy measurements demonstrate coherent and defect-free infinite-layer phase, a high structural quality with no unintentional chemical doping and a total absence of interstitial oxygen. X-ray absorption measurements show very sharp features at the Ni L$_{3,2}$-edges with a large linear dichroism, indicating the preferential hole-occupation of Ni$^{1+}$-3d$_{x^2-y^2}$ orbitals in a square planar geometry. Resonant inelastic X-ray scattering measurements reveal sharp magnon excitations of 200\,meV energy at magnetic Brillouin zone boundary, highly resonant at the Ni$^{1+}$ absorption peak. 
The results indicate that, when properly stabilized, infinite-layer nickelate thin films are superconducting without chemical doping.
}


\maketitle
\section{Introduction}
Superconductivity in Infinite-Layer (IL) nickelate thin films is under scrutiny since its discovery \cite{Li2019}, and several crucial questions remain still unanswered. Contrarily to cuprates, IL nickelates do not show a long range antiferromagnetic (AFM) order nor bulk superconductivity \cite{AbsenceBulk}, which suggests that a template is strictly necessary to properly stabilize the IL phase. The latter is obtained by following a delicate synthesis process, which poses severe problems of reproducibility \cite{Lee2020}. Because of the difficulties in the optimization of the growth and reduction parameters, the electronic and magnetic properties of these new compounds are not yet full established. From the synthesis point of view, the presence of a capping layer emerged as a relevant step to reduce the defect density, although superconductivity was demonstrated in both capped \cite{Li2019} and uncapped \cite{Zeng2020} chemically doped IL films. In particular, the absence of a capping layer largely limits the quality of the samples \cite{Raji,Parzyck2024}. Uncapped NdNiO$_2$ IL nickelate are only partially reduced, and show a (303)-stripe-organized interstitial oxygen pattern, which can be linked to the origin of the early observed charge ordering \cite{Ghiringhelli2024}. 
Similarly, the bad-metal low temperature resistance upturn observed in undoped IL nickelates, attributed to strong electron correlation effects \cite{Lee2023}, might not be an intrinsic property of the IL phase. Indeed, NdNiO$_2$ thin films realized by in-situ oxygen de-intercalation by atomic hydrogen did not show a pronounced low temperature resistance upturn \cite{Parzyck2024,Parzyck2024Synthesis}.          
\newline So far, superconductivity in IL nickelates was reported by several groups only in chemically doped samples, regardless of the approaches used to stabilize the IL phase, including the solid state Al-reduction method used in the case of Nd$_{1-x}$Eu$_x$NiO$_3$ thin films \cite{redoxAhn}, or the atomic-hydrogen method in Nd$_{1-x}$Sr$_x$NiO$_2$ \cite{Parzyck2024} and La$_{1-x}$Sr$_x$NiO$_2$ thin films \cite{Sun2024}. In this context, it is worth mentioning that superconductivity was first demonstrated in La$_{0.8}$Sr$_{0.2}$NiO$_2$ only after a significant optimization of the CaH$_2$-based process by Osada $et$ $al.$\cite{LNOsupra}. In the same report, hints of a superconducting transition, at lower temperature, were provided also for undoped LaNiO$_2$ thin films. So far, this result is not yet reproduced in other laboratories. Only very recently, an incomplete superconducting transition has been reported in the case of the undoped NdNiO$_2$ films as well \cite{parzyckSCundoped}. The observations of superconductivity in undoped IL nickelates, while needing independent verification, suggest that the reported phase diagram of nickelates, also based on the notion that undoped compounds are weakly-insulating bad-metals, \cite{PhaseDiagramLi2020,PhaseDiagramZeng2020,PhaseDiagramOsada2020,PhaseDiagramZeng2022LNO}, might need a revision. 

Here, we show that highly crystalline undoped PrNiO$_2$ thin films are superconducting, with onset transition temperature (T$_C$) in the 7-11\,K range, and a zero resistance temperature up to 4\,K. The highly reproducible zero resistance state is obtained via consecutive CaH$_2$-based topotactic reduction steps for samples prepared with a STO-capping-layer larger than 6 unit-cells (uc), while all PrNiO$_2$ films, irrespective of the presence/absence of a capping layer and/or the number of reduction cycles, show the onset of a superconducting transition. 
Scanning Transmission Electron Microscopy (STEM) shows very limited amount of defects and/or spurious phases in the entire observed volume, while divergence of the Center of Mass (dCOM) four-dimensional (4D)-STEM measurements reveals a complete absence of apical oxygens and NiO$_2$ planes much more ordered than those of NdNiO$_2$ samples \cite{krieger2024}. O K-edge electron energy loss spectroscopy (EELS) further support the absence of any unintentional chemical doping. X-ray Absorption Spectroscopy (XAS) at the Ni L$_{3,2}$-edges show very sharp peaks with a relatively large dichroism, demonstrating a complete reduction of our thin films \cite{Zeng2024}. Additionally, Resonant Inelastic X-ray Scattering (RIXS) shows that PrNiO$_2$ are characterized by well defined magnons with a bandwidth of $ca.$ 200\,meV. 
Our results suggest that the superconductivity in undoped IL nickelates might be mostly hampered by subtle details accompanying the topotactic reduction process, instead of being controlled by a threshold value of chemical doping. The data show that PrNiO$_{2}$ nickelates are self-hole-doped, as suggested by early theoretical reports \cite{Millis2020}. Finally, we show that the robust stabilization of the IL-phase is a consequence of an high-quality growth of the perovskite-phase, which is an indispensable requisite for superconductivity in undoped infinite-layer nickelates.

\section{Results}
\begin{figure}[b!]
\centering
\includegraphics[width=0.6\textwidth]{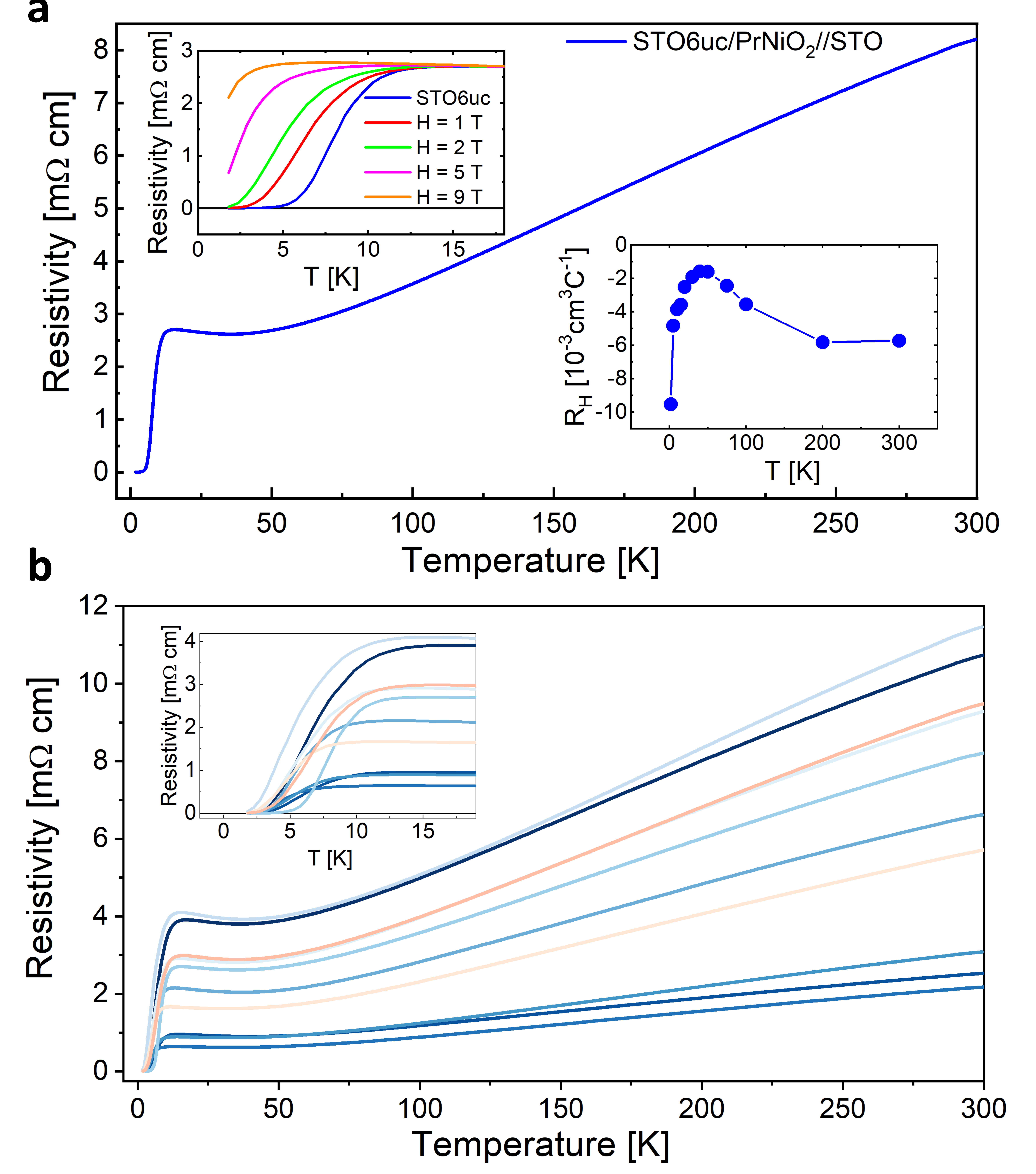}
\caption{
(a) Temperature-dependent resistivity of one of the STO(6uc)-capped, 16-unit-cell thick PNO2 film. Upper left inset shows the temperature dependence of the resistivity under applied magnetic fields; lower right inset displays the temperature-dependent Hall coefficient. (b) Temperature-dependent resistivity of several superconducting films grown in nominal similar conditions. The inset shows the same data around the superconducting transition. To  note that samples with a lower residual resistance have a much broader superconducting transition.}
\label{fig1}
\end{figure}

Precursor perovskite PrNiO$_3$ (PNO3) films were deposited onto SrTiO$_3$ (STO) single crystal as substrate and capped by an epitaxial STO film grown in-situ (See Methods and Supporting Information). A correct cation stoichiometry is mandatory to reduce extended defects and to achieve bulk-like transport properties, while granting a Ni$^{3+}$ valence state \cite{Preziosi2017, Breckenfeld2014}. It is well known that a large epitaxial mismatch can lower the energetic barrier for defects formation, that in the case of nickelates are usually identified as Ruddlesden–Popper (RP)-like stacking faults \cite{Guo2020}. These type of defects largely hampered reproducible superconductivity in Nd-based IL nickelate thin films, due to the lower crystallinity of the tensile strained perovskite precursor \cite{CationStoichiometryLi2021}. According to the perovskite nickelates phase diagram, bulk PNO3 exhibits a less distorted unit cell akin to a larger ionic radius (r$_{Pr^{3+}}$ = 0.111\,nm) compared to Nd (r$_{Nd^{3+}}$ = 0.110\,nm) \cite{phaseDiagramNickelatesPV}. The increased Ni$3d$-O$2p$ orbital overlap leads to a metal-to-insulator transition (MIT) at a relatively lower temperature ($T_{MIT}$ = 130\,K), making the material more metallic compared to NdNiO$_3$ ($T_{MIT}$ = 200\,K). The room temperature orthorhombic lattice parameters of bulk PNO3 are $a$ = 0.541\,nm, $b$ = 0.538\,nm, and $c$ = 0.762\,nm, corresponding to a pseudo-cubic lattice with $a_{pc}$ = 0.382\,nm approximately \cite{bulkPNO3}. On the other hand, the PrNiO$_2$ (PNO2) larger volume of the tetragonal (P4/mmm) unit cell ($a$ = 0.394\,nm and $c$ = 0.328\,nm) \cite{bulkPNO2} results in a higher compressive strain of PNO2 thin films onto STO (-0.9\%) compared to NdNiO$_2$ (-0.5\%). This should not be seen as a limiting factor for the stabilization of the IL phase, as there are reports of superconductivity in Nd$_{0.8}$Sr$_{0.2}$NiO$_2$ \cite{NSNO-LSAT} and Pr$_{0.8}$Sr$_{0.2}$NiO$_2$ \cite{PSNO-LSAT} thin films grown onto (LaAlO$_3$)$_{0.3}$(Sr$_2$TaAlO$_6$)$_{0.7}$ (LSAT), with hints of an enhancement of the Tc by strain. Furthermore, calculations based on a dynamical vertex approximation technique predict that, under hydrostatic pressure, the PNO2 parent compound could become a high-temperature superconductor by experiencing enhanced self-doping of the Ni orbitals \cite{PNO-pressure}.
We have used several experimental techniques to study the structural, electronic and magnetic properties of our fully reduced PNO2 samples. 
Figure~\ref{fig1}a displays the temperature dependence of the resistivity for 
one of our first superconducting PNO2 film, composed of 16 unit-cells (uc)-thick PNO2 film capped with 6 uc of STO (hereafter referred as STO6uc-PNO2). Despite a relatively large resistivity at room temperature, compared to other reports in literature \cite{Osada2020, araceli2024}, the metallic T-linear behavior is followed by a slight upturn below $ca.$ 50\,K and a complete superconducting transition around 4\,K. The measured onset critical temperature T$_C$ (maximum curvature) is slightly below 11\,K. This represents the main result of this work. In the top-left inset of Figure~\ref{fig1}a we show the temperature-dependence of the resistivity as a function of the perpendicularly applied magnetic field, up to 9 Tesla. The normal state resistivity is insensitive to the magnetic field, while superconductivity is progressively suppressed. In the bottom-right inset of Figure~\ref{fig1}a we show the temperature dependence of Hall coefficient, which is negative in the whole temperature range. The Hall coefficient shows a tendency towards a change of sign down to 50\,K, and then an opposite trend. Nearby T$_C$-onset, its negative value further increases, which is remarkably different from change of sign, from negative to positive, observed in superconducting chemically doped samples \cite{Li2019,PhaseDiagramOsada2020,Ariando2022}. On the other hand, the Hall coefficient temperature dependence is very similar to that one reported on undoped (non-superconducting) samples.

In Figure~\ref{fig1}b we show the transport properties of a series of STO6uc-PNO2 samples prepared in nominal similar conditions. All the samples display a zero resistance state below 4\,K. The T$_C$-onset varies from sample-to-sample in a relatively small range, with a minimum value of 7\,K and maximum value of 11\,K, while the normal state residual resistivity and its value at room temperature  varies in a much larger range, likely reflecting subtle details in the topotactic reduction process. The superconductivity in these undoped PNO2 films is very robust. In particular, an incomplete superconducting transition, with similar T$_C$-onset, is observed also in uncapped PNO2 samples, although  a zero resistance state is obtained only on PNO2 capped with at least 6uc of STO (See Supplemental Information for additional data). This confirms the key role played by the capping layer in stabilizing a clean and robust IL phase. 

\begin{figure}[b!]
\centering
\includegraphics[width=0.9\textwidth]{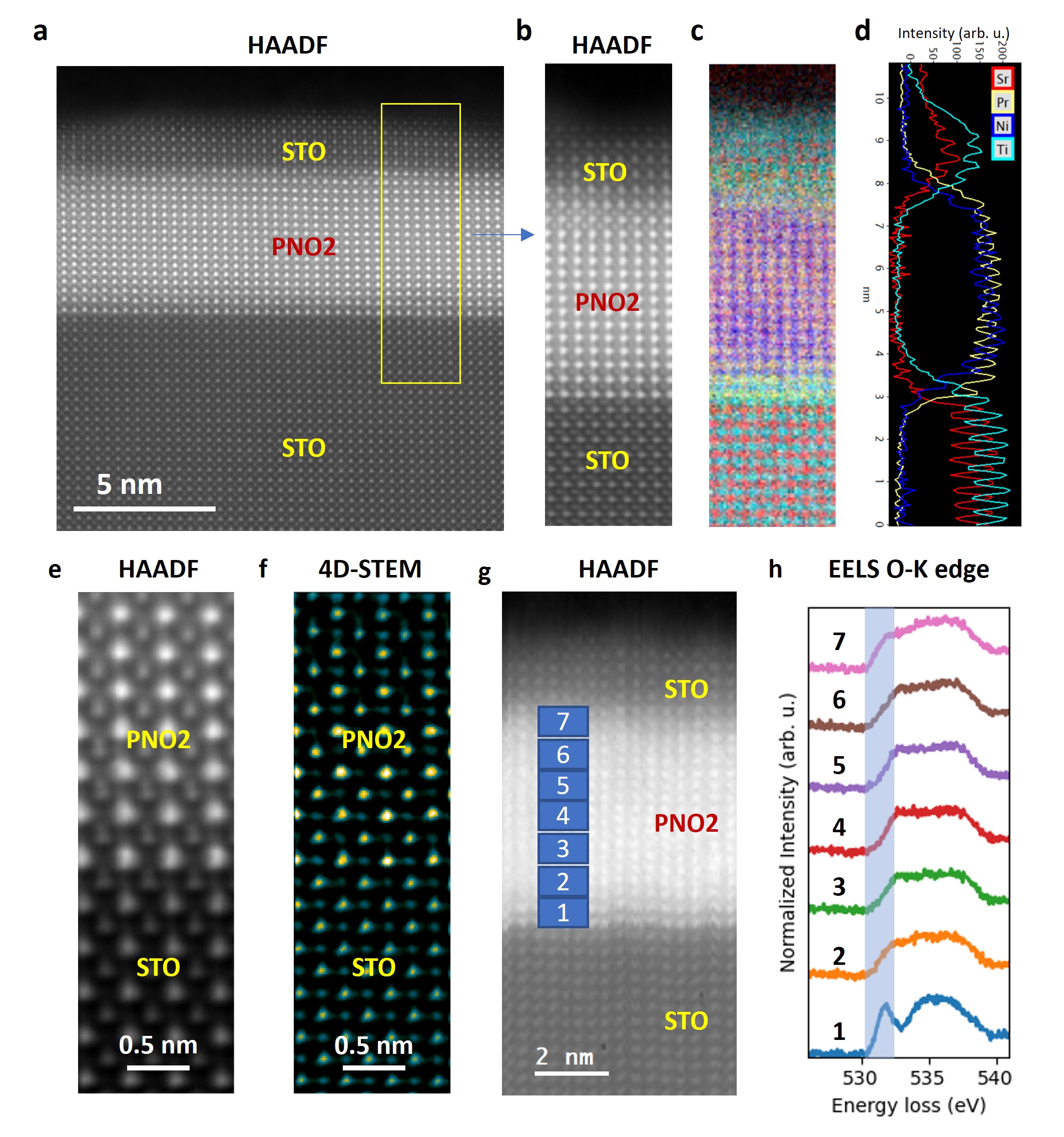}
\caption{(a) HAADF-STEM image of a 6-unit-cell STO capped PNO2 film (16uc thick). (b) HAADF image from a zoom-in region. (c) EELS Sr, Pr, Ni, and Ti elemental maps, and (d) elemental profiles integrated over the zoom-in region showing a Pr/Ni ratio close to 1 over the whole thin film. Both top and bottom interfaces are not sharp. (e) A magnified HAADF image from the bottom interface region, (f) the corresponding 4D-STEM dCOM analysis showing clean infinite-layer phase without the presence of any apical oxygen. (h) Real space evolution of  the O-K edge EELS fine structure from regions as labelled in the HAADF image in (g).}\label{fig2}
\end{figure}

We resorted to STEM measurements with a High-Angle Annular Dark-Field (HAADF) imaging technique and Electron-Energy-Loss Spectroscopy (EELS) to study the precise atomic stack, including both bottom and top interfaces. The atomic-resolved HAADF-STEM cross-sectional image shown in Figure~\ref{fig2}a displays a clear infinite-layer phase with a very high structural quality, thus demonstrating the overall absence of RP-like defects in the precursor phase, as confirmed by HAADF-STEM cross-sectional images for STO6uc-PNO3 samples. 
The EELS data acquired over the entire STO/PNO2//STO stack of a selected area of the HAADF image (Fig. \ref{fig2}b), show a clear intermixing at the bottom film-substrate and at the top film-STO interfaces (Figure~\ref{fig2}c). In particular, the interface with the TiO$_2$-terminated substrate is characterized by a one unit cell (Ni,Ti) intermixing layer, confirming other reports \cite{Goodge2023,Raji2024}.
Additionally, the total element map intensity profile shown in Figure~\ref{fig2}d clearly indicates that the top interface exhibits a complex Ni-Pr-Ti-Sr atomic stack, which differs from other recent reports of more abrupt interfaces in Sr-doped PNO3 and related IL thin films \cite{Raji2024}. It is worth to recall here that, early theoretical calculations proposed the formation of  a quasi-2D electron gas, with possible superconducting properties, in the case of a clean interface between the infinite-layer nickelate and the STO substrate \cite{Theory2DEG}. Our HAADF-STEM results, in agreement with Goodge $et$ $al$., fully rules out the formation of a quasi-2D electron gas since the intermixed (Ni,Ti)O$_3$ interface-layer fully compensates the polar discontinuity \cite{Goodge2023}.

Unintentional chemical doping can arise from Pr/Ni off stoichiometry, Sr-diffusion and interstitial/apical oxygen incorporated into the nickelate layer.
Our elemental EELS mapping shows a near unitary Pr/Ni ratio and no unintentional Sr-doping (within the experimental resolution of the order of 10\%) in the PNO2 sample. 
In order to identify any traces of these interstitial/apical oxygen, we employed 4D-STEM dCOM imaging. Some of the authors have already shown that this powerful technique allows high-resolution real space atomic mapping with good oxygen contrast \cite{Raji}. The HAADF image in Figure~\ref{fig2}e, is obtained at the bottom interface, and the corresponding 4D-STEM dCOM analysis in Figure~\ref{fig2}f unveils the occupied oxygen sites not visible in the HAADF image. Excluding the intermixed interfacial unit-cell, characterized by a (Ni,Ti)Ox composition in an octahedral coordination, we can easily distinguish occupied oxygen-sites only within the NiO$_2$ planes as, indeed, expected for a properly stabilized IL-phase. This directly excludes traces of apical oxygen within the PNO2 samples and, moreover, demonstrates a rather clean Ni-sites in square planar arrangement. 
To further evaluate the absence of any chemical source of doping with higher sensitivity, we performed atomically resolved EELS fine structure analysis at the Oxygen K-edge. This is shown in Figures \ref{fig2}g,h. Figure~\ref{fig2}h shows spatially resolved STEM-EELS of O K-edge fine structures in different regions of the sample, including the two interfacial regions. As expected for fully reduced IL \cite{Goodgee2007683118,Raji}, the O K-edge profiles do not show any pre-peak feature, with the only exception of the area in proximity of the substrate, where a certain degree of Ni$3d$-O$2p$ hybridization is still observed because of the (Ni,Ti) intermixing layer \cite{Bisogni2016}, and nearby the top, intermixed, interface due to the STO-capping-layer. The absence of any pre-edge is a clear indication that no unintentional chemical doping is present in our samples. In particular, a pre-peak shoulder is always observed, also locally, in Sr-doped IL-nickelates, as shown in previous works \cite{Rossi2020,Goodgee2007683118,Raji2024}. From this analysis, we can confidently conclude that the residual chemical doping is below the sensitivity of the multiple techniques used, and thus much lower than the minimum threshold needed so-far to trigger superconductivity in chemically doped IL-nickelates.

\begin{figure}[t!]
\centering
\includegraphics[width=1\textwidth]{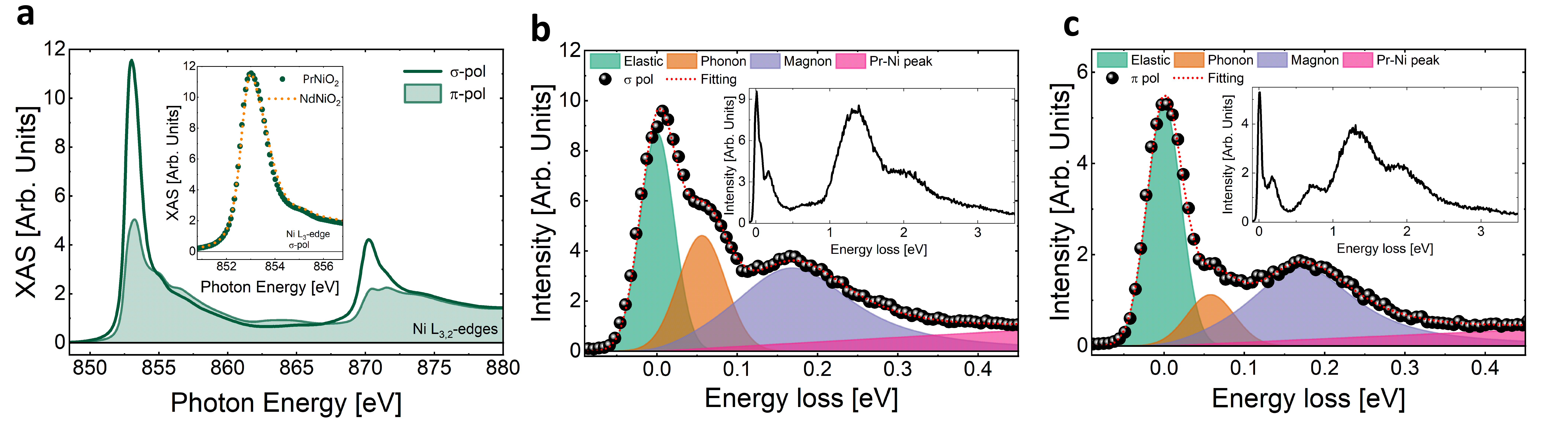}
\caption{(a) Linearly polarized XAS spectra at the Ni L$_{3,2}$ edges for a STO6uc-PNO2 sample, with electric field parallel ($\sigma$-pol) and perpendicular ($\pi$-pol) to the NiO$_2$ planes. The inset shows the comparison with a XAS acquired in $\sigma$-pol at the Ni L$_{3}$-edge for a NdNiO$_2$ sample. (b) Mid-infrared region of the $\sigma$-pol RIXS spectrum along the transferred momentum $\bf{Q}$ (-0.36,0). The spectrum was obtained as a sum of three single spectra, with incident energies 852.3\,eV, 852.5\,eV and 853.7\,eV, to reduce the noise. The full-range spectrum is reported in the inset. (c) Mid-infrared region of the $\pi$-pol RIXS spectrum acquired at 852.5\,eV photon energy. The spectrum was obtained as a sum of four single spectra, with exchange momentum H = 0.4, 0.425, 0.45, 0.475 r.l.u. along the [H,0] direction of the Brillouin zone, to reduce the noise. The shaded colorful areas in panels b-c are the results of a fitting procedure as presented in Methods. All the measurements were performed at T = 20\,K $ca.$}
\label{fig3}
\end{figure}

In order to get information about the electronic structure of our superconducting PNO2 thin films, we performed XAS and RIXS measurements at the Ni L$_{3,2}$-edges and L$_3$-edge, respectively (see Methods for details). Overall, the Ni L$_3$ XAS spectra show features similar to properly optimized NdNiO$_2$ thin films \cite{krieger2024,Rossi2020,Zeng2024} with a dominant peak due to the 2p$^6$3d$^9$ $\rightarrow$ 2p$^5$3d$^{10}$ transition. Figure~\ref{fig3}a shows XAS spectra acquired with linearly-polarized light in the parallel and nearly perpendicular direction with respect to the NiO$_2$ planes, $i.e.$ $\sigma$-pol and $\pi$-pol, respectively. The more than 50\% dichroism is a direct consequence of the very robust infinite-layer phase, where the majority of the holes resides in the Ni$^{1+}$-3d$_{x^2-y^2}$ orbitals. These results confirm a full oxygen reduction after the topotactic process. Moreover, the Ni L$_3$ peak is very sharp with a small shoulder at higher photon energy, speaking against adventitious sources of doping due to incomplete reduction (Ni$^{2+}$) and/or excess oxygen \cite{Rossi2020,krieger2024}. For comparison, the inset in Figure~\ref{fig3}a reports the XAS at the Ni L$_3$-edge peak acquired for a STO-capped NdNiO$_2$ with, indeed, very similar features.   
In Figures \ref{fig3}b,c we show the mid-infrared region of RIXS spectra acquired in $\sigma$-pol and $\pi$-pol, respectively. We can easily recognize both the phonon and magnon excitations, which we fit with a Gaussian and a Damped Harmonic Oscillator susceptibility, respectively (please refer to the Supporting information for further details about the fitting procedure). A linear background from the tails of higher-energy excitation is also included. The phonon is apparently enhanced by $\sigma$ incident polarization and, therefore, mostly associated to in-plane Ni-O stretching modes, similarly to cuprates \cite{braicovich2020determining}. The magnetic peak occupies an energy range around $ca.$ 200\,meV, consistently with previous results on IL nickelates \cite{Gao2024,rosa2024spin,Lu213,fan2024capping}. In the insets we show the same RIXS spectra in the full energy loss range, enough to capture the Ni 3d orbital excitations betweeen 1 and 3\,eV \cite{Rossi2020}. The peak at 0.6\,eV, which is more pronounced in $\pi$-pol, is attributed to out-of-plane Pr-Ni hybridization, in analogy with other undoped nickelates \cite{Hepting2020,Rossi2020,KriegerPRL,rosa2024spin}. Finally, the shape of the magnetic excitations, among the sharpest ever measured by RIXS in undoped IL nickelates, though not resolution-limited, indicates that superconductivity is reached in the presence of long-range spin-spin correlations. This means that either a self doping mechanism is present, which does not perturb the antiferromagnetic background, or that superconductivity is obtained at extremely low levels of hole doping.
\section{Discussion}
The main finding of this study is the observation of a superconducting ground state in undoped PrNiO$_2$ thin films: below, we briefly discuss our finding and its implications.
The synthesis of infinite-layer nickelates remains a real challenge, and one has to make sure that the reported superconductivity in nickelates parent compounds PNO2 thin films is not due to fortuitous sources of doping. There are at least three sources of possible unintentional chemical doping: adventitious Sr-doping; Pr/Ni off-stoichiometry; and excess oxygen due to incomplete de-intercalation during the CaH$_2$-based topotactic reduction.
The combination of STEM-EELS, layer resolved O K-edge EELS, and 4D-STEM analysis on PNO2 samples in Figure~\ref{fig2}, largely discussed in the previous section, clearly demonstrate that, locally, no relevant chemical-doping could be detected. Moreover, XAS spectra ($cf.$ to Fig. \ref{fig3}a) show very sharp Ni$^{1+}$ peaks, at odds with possible doping by chemical substitution, which will give signatures of Ni$^{2+}$ features \cite{krieger2024}. 

\begin{figure}[t!]
\centering
\includegraphics[width=0.7\textwidth]{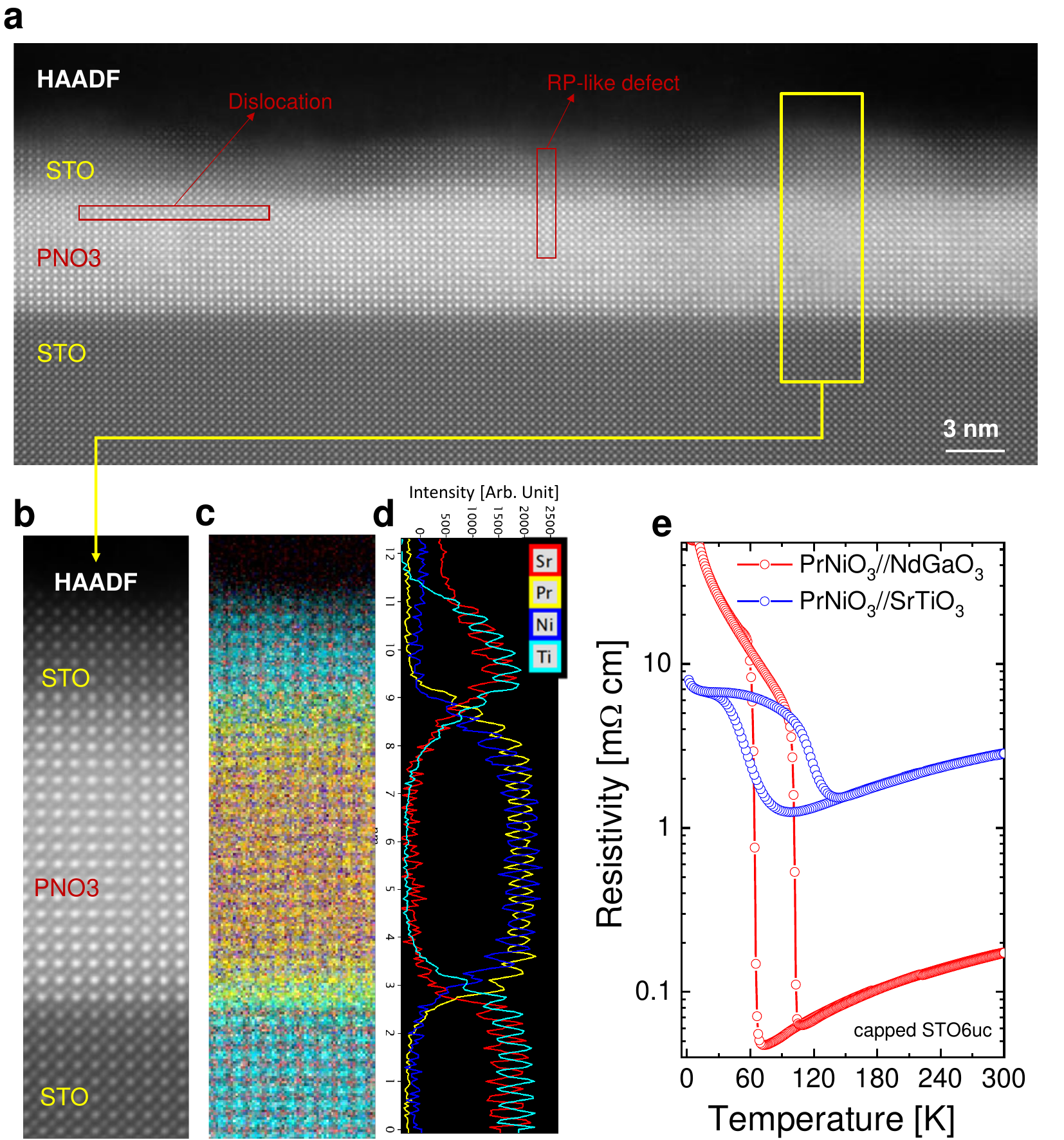}
\caption{(a) HAADF-STEM image of the STO6uc-PNO3 film. (b) A defect-free zoom-in region. (c) Elemental STEM-EELS maps of Sr, Pr, Ni, and Ti. (d) Elemental profiles integrated over the zoom-in region showing a Pr/Ni ratio close to one throughout the investigated sample volume. Importantly, no Sr-interdiffusion could be resolved. (e) Temperature-dependent resistivity of 16 unit-cells thick PNO3 film grown onto STO and NdGaO$_3$ substrates, and both capped with 6 unit-cells of STO.}
\label{fig4}
\end{figure}

Here now we discuss the relevance of the structural and chemical perfection of the perovskite precursor phase, which allows to provide further support of the absence of a relevant unintentional chemical doping related to cation off stoichiometry. From the STEM-EELS map on PNO2 in Figure~\ref{fig2}, we found that the Pr/Ni ratio is close to one. However, to further exclude Pr/Ni off-stoichiometry, in Figures \ref{fig4}a-d we report atomic-resolved HAADF-STEM (Figs. \ref{fig4}a,b) and EELS maps (Figs. \ref{fig4}c,d) of our STO6uc-PNO3 precursor samples. We find that the Pr/Ni ratio is equal to one within the experimental uncertainty, thus further excluding off-stoichiometry issues. Additionally, in Figure~\ref{fig4}e we report the transport properties of STO6uc-PNO3 precursor samples deposited onto STO, and on well-matched NdGaO$_3$ (110) single crystal grown in the same deposition conditions, thus characterized by the same Pr/Ni composition \cite{BenckiserPRM2021}. In these nickelates, the sharpness of the metal-to-insulator transition (MIT) is a valid proxy to gain major information about the Pr/Ni ratio \cite{Preziosi2017}. The data show a MIT with a reduced jump in the case of the PNO3//STO samples due to strain but still a very high MIT temperature, in agreement with other studies \cite{PhaseDiagramOsada2020}, while a very large resistance jump for the PNO3//NGO film, comparable to PrNiO$_3$ single crystals \cite{Saito2003}. It is worth noting that the observation of a MIT in our precursor phase is a further indication of the absence of any Pr/Ni off-stoichiometry and even Sr-doping. Indeed, it is well known that the MIT is suppressed by a small amount of Pr/Ni off-stoichiometry as well as by few percent of Sr-doping. All these results imply a nearly unitary Pr/Ni ratio in our films, ruling out non-unitary Pr/Ni ratio as a source of unintentional doping, and also Sr-doping.

Beyond material imperfections, superconductivity in undoped nickelates, on the other hand, might be related to self-doping of holes due to the R$5d$-Ni$3d$ hybridization. According to theoretical calculations, the R$5d$ states gives rise to partially occupied electron pockets at the $\Gamma$-point, and the hybridization with Ni-3d$z^2$ states at the Fermi level provides self-doped holes in the the quasi-2D 3d$x^2-y^2$ band \cite{Millis2020}. This specific electronic configuration, while setting one of the major differences with the cuprates' fermiology, may be responsible of a superconducting state already at zero chemical doping \cite{Pickett2020}.

To summarize, we report clear experimental evidences that undoped PrNiO$_2$ IL-nickelates are superconducting, at odds with the observation of superconductivity only in IL doped by Sr,Ca cations, which provide extra-holes in the NiO$_2$ planes \cite{Li2019,Lee2020,Ariando2022,PhaseDiagramOsada2020}.
Our results, together with the report of the onset of superconductivity in undoped LaNiO$_2$ thin films \cite{LNOsupra}, and more recently in undoped NdNiO$_2$ thin films \cite{parzyckSCundoped}, represent an important breakthrough for a full understanding of infinite-layer nickelates physics. According to these studies, due to the multi-orbital nature of the electronic properties of nickelates, self-doped holes in the Ni-3d$x^2-y^2$ orbital-derived bands, are enough to set a superconducting state. This requires a reconsideration of the IL nickelate phase-diagram.

\section*{Methods}\label{Methods}
\paragraph{Thin Film Growth:}
The epitaxial growth of perovskite PrNiO$_3$ films on $5\times5 mm^2$ SrTiO$_3$ substrates was performed by pulsed laser deposition using a 248 nm KrF excimer laser, with ceramic targets from Toshima Manufacturing Co. Ltd. 
The SrTiO$_3$ substrates (Shinkosha Co. Ltd) were prepared by etching in an NH$_4$F-buffered HF solution, followed by annealing for 2 hours at 950$^\circ$C in air to achieve a well-defined TiO$_2$-terminated step-terraced surface. 
Prior to growth, the substrate was pre-annealed for 1 hour at 800$^\circ$C under a pressure of 0.3 mbar in oxygen flow to ensure a very clean and sharp step-and-terrace surface. 
The layer-by-layer growth of the PrNiO$_3$ was monitored via Reflection High Energy Electron Diffraction (RHEED) technique. The 15-20 unit cells thick PrNiO$_3$ films were grown at a substrate temperature of 675$^\circ$C with an oxygen partial pressure P$_{O_2}$ of 0.3 mbar, using a laser fluence of 4 J/cm$^2$ and a $1 \times 1.4$ mm$^2$ laser spot size on the target.For capped samples, the SrTiO$_3$ top layers (from 1 to 12 unit cell thickness), were grown at a substrate temperature of 575$^\circ$C and P$_{O_2}$ = 0.3 mbar, with a laser fluence of 1.3 J/cm$^2$ and a $1 \times 1.4$ mm$^2$ laser spot size.
After growth, the samples were cooled to room temperature at a rate of 5$^\circ$C/min in the same oxidizing growth conditions.
\paragraph{Topochemical Reduction (infinite-layer formation):}
After growth, each sample was cut into two pieces of size $5 \times 2.5$ mm$^2$ using a precision diamond wire saw (Well 3242). This method is used to cut the samples to avoid stress during the cutting process, which can introduce defects. 
The pieces of each sample to be reduced were placed in an evacuated silica tube sealed with a membrane valve, with 0.5\,g of CaH$_2$ powder in direct contact, as used in prior studies \cite{Krieger2023}. 
The tube was heated to 260$^\circ$C at a rate of 5$^\circ$C/min, and held at this temperature for varying durations (2-8 hours) depending on the PrNiO$_3$ and SrTiO$_3$ capping thicknesses; then it was cooled to room temperature at a rate of 5$^\circ$C/min. 
The process was optimized through a systematic series of steps, with $\theta$-$2\theta$ scans and electrical transport measurements performed ex situ after each step to assess the degree of reduction. In particular, the highly reproducible zero resistance state is obtained via consecutive CaH$_2$-based topotactic reduction steps for samples prepared with a STO capping layer (from 3 up to 12 unit-cells), while a superconducting transition is always encountered irrespective of the presence/absence of the capping layer and/or number of reduction cycles.
\paragraph{Characterization:}
The surface morphology of the samples was examined using a Park XE7 (Park System) atomic force microscope (AFM) in true non-contact mode. 
XRD measurements were performed using a Rigaku Smartlab diffractometer with a Cu-K$\alpha$ radiation source (0.154056\,nm). For transport measurements, the 2.5$\times$5 mm$^2$ samples were wire-bonded directly with Al, without the use of top-electrodes. 
The wire bonds were placed at the four edges of the samples, and the measurements were conducted using the Van der Pauw method, with a current amplitude of 10 $\mu$A, with a cryo-free Dynacool system (Quantum Design).
\paragraph{High resolution STEM-EELS:}
The cross-sectional focused ion beam (FIB) transmission electron microscopy (TEM) lamellae were prepared at C2N, University of Paris-Saclay, France. Before FIB lamellae preparation, around 20\,nm of amorphous carbon was deposited on top for protection. For additional protection, electron beam-induced deposition of platinum and ion-beam beam-induced deposition of platinum were done. The HAADF imaging and EELS were carried out in a NION UltraSTEM 200 C3/C5-corrected STEM. The experiments were done at 200\,keV with a probe current of approximately 14\,pA and convergence semi-angles of 30\,mrad. The EELS spectra were obtained using the full 4 × 1 configuration of a MerlinEM detector (Quantum Detectors Ltd) installed on a Gatan ENFINA spectrometer mounted on the microscope \cite{Blank2014}. The EELS spectrometer was set into non-energy dispersive trajectories for 4D-STEM experiments. Data was collected with only one chip of the MerlinEM detector in 6-bit mode that enables faster acquisition without compromising on the signal-to-noise ratio. The high resolution EELS fine structure analysis was done in a monochromated and C3/C5 corrected NION Chromatem microscope operating at 100 keV, with a probe current of 30 pA, convergence semi-angles of 25 mrad.  
\paragraph{XAS, RIXS:} 
Measurements were performed at the ID32 soft x-ray beamline of the ESRF, Grenoble, France. XAS spectra were measured at 10$^{\circ}$ grazing incidence, with linear polarization forming an angle of 0$^{\circ}$ or 80$^{\circ}$ with respect to the NiO$_2$ planes, and labelled for brevity $\parallel$NiO$_2$ and $\perp$NiO$_2$, respectively. For RIXS, the combined resolution at the Ni L$_3$-edge was of 40\,meV. The scattering angle 2$\theta$ was fixed at 149.5$^{\circ}$. RIXS energy-resolved maps were acquired at an incident angle of $\theta = 30^{\circ}$ (grazing-in geometry), corresponding to an exchanged in-plane momentum of about 0.36 relative lattice units (r.l.u.). Momentum-resolved maps were taken at the XAS resonance energy of 852.4\,eV ca. in $\pi$ incident polarization, which is known to enhance magnetic excitations. Grazing-out geometry was adopted in this case, with incident angles $\theta$ between 80$^{\circ}$ and 140$^{\circ}$. All RIXS and XAS measurements were performed at 20 K, the lowest temperature safely reachable by the ID32 cooling apparatus.

\clearpage
\appendix
\section*{\fontsize{14pt}{18pt}\selectfont Supporting Information}\label{Supporting Information}
\setcounter{figure}{0}
\renewcommand{\thefigure}{S\arabic{figure}}

\paragraph{Growth of SrTiO$_3$(d)/PrNiO$_3$//SrTiO$_3$ heterostructures}

\begin{figure}[b!]
\centering
\includegraphics[width=0.6\textwidth]{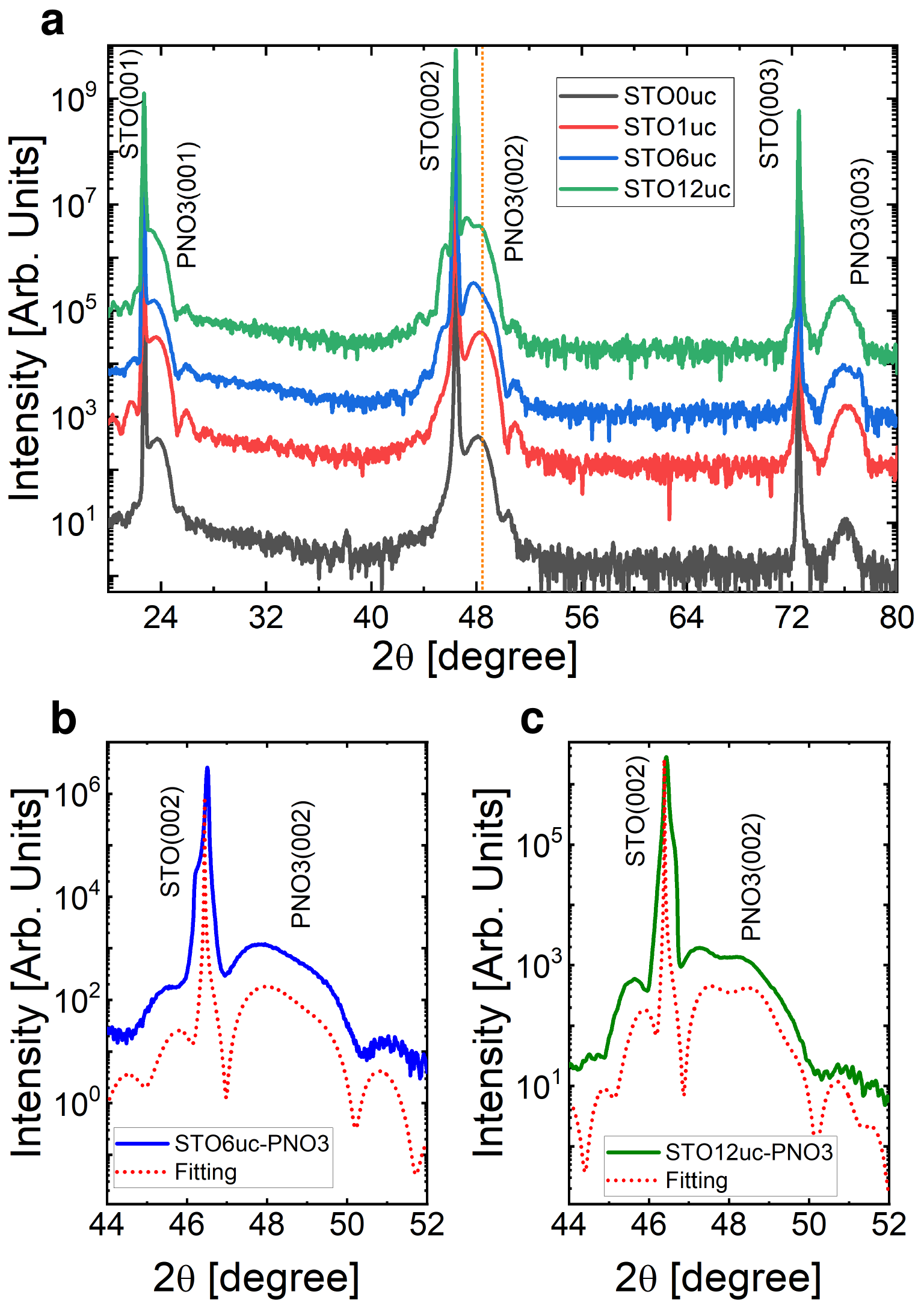}
\caption{
a) X-ray diffraction $\theta$-2$\theta$ symmetric scans of PNO3 sample series with varying thicknesses of the STO capping layer. Below we show a zoomed-in around the (002) diffraction peak showing the peculiar modulation as described in the text together the fitting curves (red dotted lines), obtained by considering the presence of an epitaxial and coherent STO layer of 6uc (b) and (c) 12 uc as thickness. The fitting curves are obtained by using the software described in Ref. \cite{FittingXRD}.
}
\label{figS1}
\end{figure}

In Figure~\ref{figS1}a, we show the XRD $\theta$-2$\theta$ symmetric scans of 16 unit cells (uc) thick PNO3//STO films capped with varying thicknesses (d) of STO, ranging from 0 to 12 uc. 
For all samples, intense (00$\ell$) diffraction peaks ($\ell$ = 1, 2, 3) are observed, confirming the high quality of the perovskite nickelate phase \cite{Krieger2023}. 
From the (00$\ell$) peak positions, we calculate a c-axis lattice parameter of 0.375\,nm, consistent with a fully strained and stoichiometric PNO3 thin film. 
Accounting for a Poisson ratio of 0.3, the expected out-of-plane lattice parameter for a tensile-strained PNO3 thin film onto STO is approximately of 0.375\,nm, thus confidently ruling out any possible presence of off-stoichiometry and/or oxygen vacancies \cite{Preziosi2017,Breckenfeld2014}. 
A zoom-in around the (002) diffraction peaks (Figures \ref{figS1}b,c) reveals an intensity modulation of the peak profile for the PNO3 sample prepared with an STO capping layer thicker than 6uc. 
The HAADF-STEM results for the STO6uc sample (Figure~\ref{fig4}b) clearly demonstrate that this intensity modulation of the (002) XRD peak profile is not due to secondary phases or hole-vacancies, but is most likely a direct consequence of the STO capping layer itself as, indeed, we carefully demonstrated by performing a fitting procedure by using a tool already used in literature \cite{FittingXRD}. The red dotted lines superimposed to the XRD pattern of our STO6uc-PNO3 and STO12uc-PNO3 samples properly reproduce the observed PNO3(002) XRD peak modulation.

\paragraph{Structural and transport properties of SrTiO$_3$(d)/PrNiO$2$//SrTiO$_3$ heterostructures}

\begin{figure}[t!]
\centering
\includegraphics[width=0.9\textwidth]{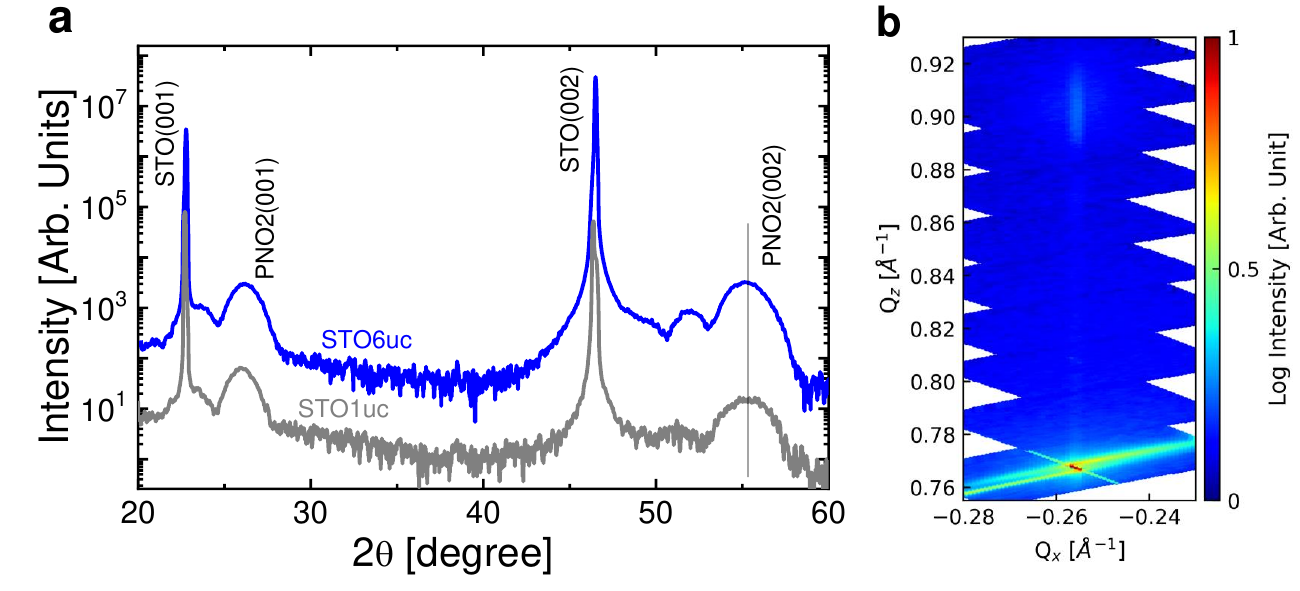}
\caption{
a) X-ray diffraction $\theta$-2$\theta$ symmetric scans of STO6uc- and STO1uc-PNO2 samples, clearly showing the formation of the infinite-layer (IL) phase. b) Reciprocal space mapping performed around the asymmetric (103) STO diffraction peak, indicating that the samples are fully strained and exhibit a 0.333\,nm c-axis parameter.
}
\label{figS2}
\end{figure}

To stabilize the PrNiO$_2$ (PNO2) infinite-layer (IL) phase, we employed a CaH$_2$-based topotactic reduction process. 
Figure~\ref{figS2}a shows the XRD $\theta$-2$\theta$ symmetric scans of fully reduced samples capped with STO of 6 unit cells (STO6uc) and 1 unit cell (STO1uc), exhibiting robust (00$\ell$) family peaks at the expected positions, with a $c$-axis of 0.333\,nm, confirming the complete formation of the IL phase. 
Notably, features on the left of the main (00$\ell$) diffraction peaks in the STO6uc-capped sample are likely attributed to Laue fringes, indicating the high quality of the IL phase, stabilized coherently by the appropriate thickness of the capping layer \cite{Raji2024,Lee2020}.
\begin{figure}[t!]
\centering
\includegraphics[width=0.65\textwidth]{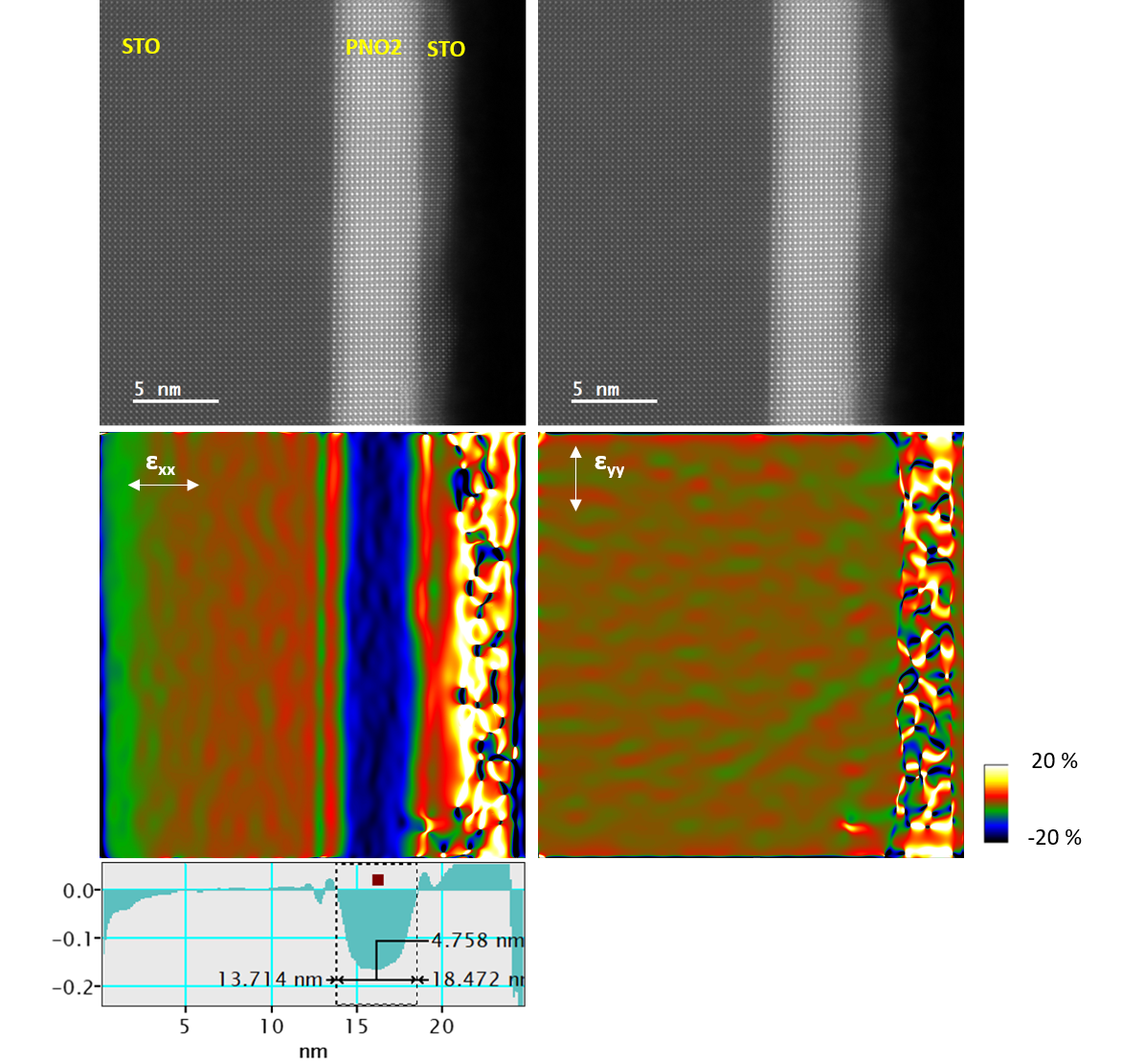}
\caption{
(Top) HAADF-STEM images for a PNO2 sample capped with 6 unit cells of STO, with (Bottom) GPA analysis applied, confirming a 15-16\% reduction in the out-of-plane c-axis and full in-plane strain of the IL-phase. This reduction confirm the overall XRD result and shown in \ref{figS2} 
}
\label{figS3}
\end{figure}
Reciprocal Space Mapping (RSM) measurements around the asymmetric STO(-103) and PNO2(-103) diffraction peaks confirm that the PNO2 films are fully strained to the substrate and show no other features, as shown in Figure~\ref{figS2}b.
This is further corroborated by a geometrical phase analysis (GPA) map obtained via HAADF-STEM measurements, shown in Figure~\ref{figS3}, which displays a homogeneous 15\% reduction in the $c$-axis (GPA-$\epsilon_{xx}$), supporting the macroscopic observations via XRD and indicating a fully strained IL-phase to the substrate (GPA-$\epsilon_{yy}$).

\begin{figure}[h]
\centering
\includegraphics[width=0.5\textwidth]{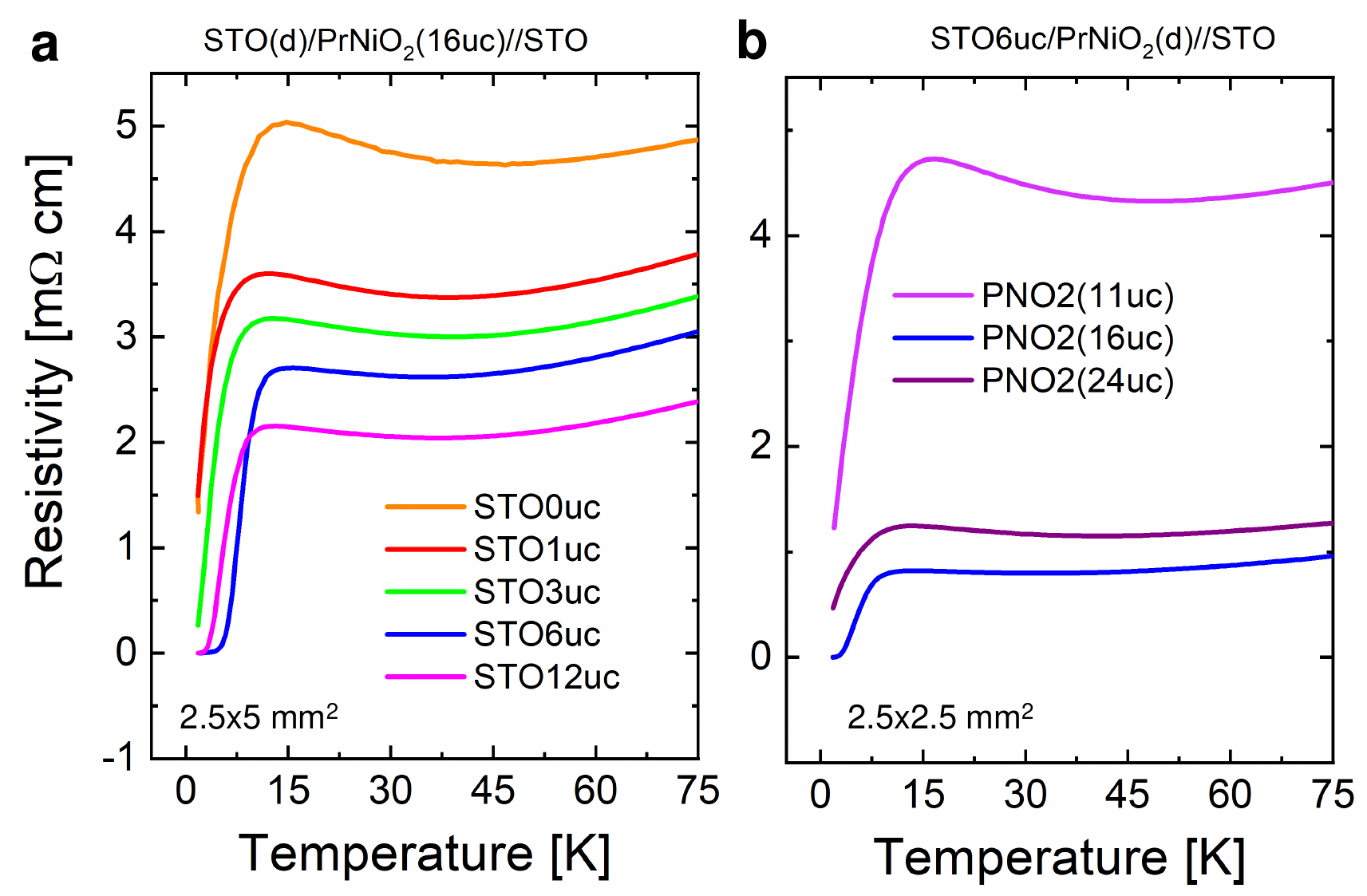}
\caption{Temperature-dependent resistivity of the superconducting PNO2 sample series, prepared with (a) varying unit cells of the STO-capping layer but constant PNO2 thickness and (b) varying unit cells of PNO2 but constant STO-capping layer. The zero resistance state is achieved for samples with moderate PNO2 thickness (16 uc) and STO layers thicker than 6 unit cells.}
\label{figS4}
\end{figure}

In Figure~\ref{figS4} we show the transport properties of the full PNO2 series prepared with different STO(d) capping layer thicknesses (left), and also a series of PNO2(d) samples with the same STO6uc capping layer and overall similar conditions for the topotactic reduction.
While the superconducting transition is always present, only samples with a capping layer thicker than 6 unit cells reach the zero resistance state. On the other side the zero-resistance state in encountered only for sample 16uc thick. This largely demonstrate that the structural quality of the precursor phase (eventually worsened for thicker sample) is a key parameter to obtain a superconducting state with a clear zero resistance state. This demonstrates, beyond the HAADF-STEM data, that the PNO2 superconductivity is not linked to any possible source of Sr interdiffusion.

\begin{figure}[h]
\centering
\includegraphics[width=0.65\textwidth]{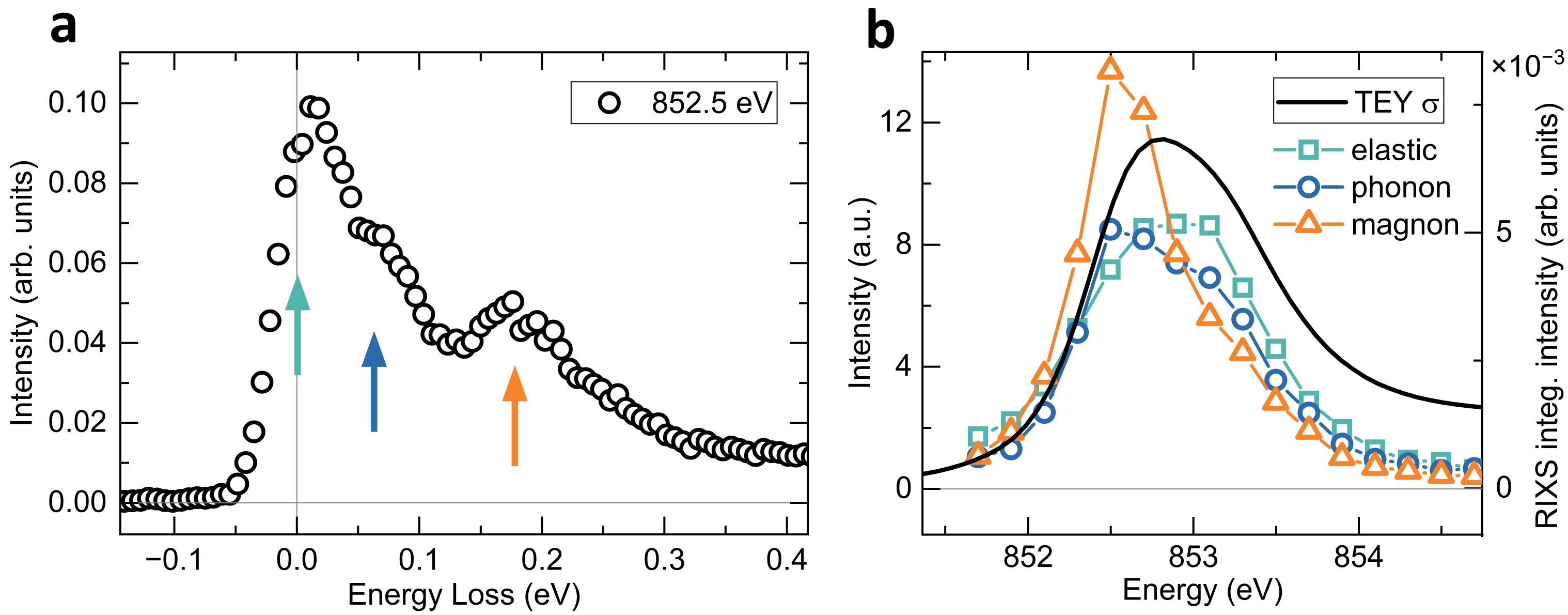}
\caption{(a) RIXS spectrum at the magnon resonance, taken with $\pi$ incident polarization. The arrows indicate the main spectral features in the low energy scale: the elastic line, the phonon peak and the magnon spectral weight. (b) Energy dependence of the different spectral features highlighted in panel (a) across the Ni L$_3$ edge. The integration intervals are [-0.03 eV, 0.03 eV], [0.03 eV, 0.1 eV] and [0.1 eV, 0.35 eV] for the elastic, phonon and magnon respectively.}\label{figS7}
\end{figure}

\paragraph{Energy dependence of the magnon}
As already suggested by the colour maps in Fig.~\ref{fig3}(b), the energy dependence of the single magnon follows a peculiar trend with respect to the other main spectral features (e.g., elastic peak, phonon). In particular, as shown in Fig.~\ref{figS7}, the magnon intensity resonates before the absorption edge, $\simeq0.4$ eV de-tuned from the XAS peak, as opposed to the elastic and the phonon intensity. Moreover, the magnon resonance appears to be significantly narrower than the other two features presented in Fig.~\ref{figS7}(b).

\paragraph{Fitting of RIXS spectra}

Here we introduce the fitting procedure for the results for the RIXS spectra showed in Figure~\ref{fig3}c of the main text. A Damped Harmonic Oscillator susceptibility is employed for the magnetic peak, while two resolution-wide Gaussians fit the elastic and phonon contribution. The tail of the broad 0.6\,eV peak due to Pr$5d$ hybridization is taken into account with a linear background. Fittings close to the Gamma point are complicated by the softening magnon, which tends to merge with other low-energy features. This is testified by the large error bars on the damping at such values of Q. Nevertheless, our fitting results are in good agreement with literature \cite{Lu213,rossi2024universal}, including our very recent measurements on NdNiO$_2$ \cite{rosa2024spin}.

\bmhead{Acknowledgments}
The authors thank enriching discussions with K.M. Shen. This work was funded by the French National Research Agency (ANR) through the ANR-JCJC FOXIES ANR-21-CE08-0021. This work was also done as part of the Interdisciplinary Thematic Institute QMat, ITI 2021 2028 program of the University of Strasbourg, CNRS and Inserm, and supported by IdEx Unistra (ANR 10 IDEX 0002), and by SFRI STRAT’US project (ANR 20 SFRI 0012) and EUR QMAT ANR-17-EURE-0024 under the framework of the French Investments for the Future Program. The authors would like to thank the PLD, XRD, MEB-CRO, TEM and MagTransCS platforms of the IPCMS. The RIXS experiments were performed at ESRF Synchrotron facility in Grenoble (France) under proposal number SC-5438. The ID32 staff is also acknowledged for technical support.
\clearpage
\bibliography{biblio}

\end{document}